\documentclass[12pt]{article}
\usepackage{amsmath}
\usepackage[dvips]{graphicx}
\usepackage{epsfig}
\usepackage{amsmath}
\usepackage{amssymb}
\usepackage{graphics,epstopdf}
\usepackage{amsfonts}
\usepackage{amsthm}
\usepackage[usenames]{color}

\definecolor{AV}{rgb}{0.65,0.0,0}
\definecolor{GC}{rgb}{0,0.0,0.65}
\definecolor{WS}{rgb}{0,0.65,0}

\usepackage[
      colorlinks=true,
      linkcolor=blue,
      urlcolor=blue,
      filecolor=blue,
      citecolor=red,
      pdfstartview=FitV,
      pdftitle={},
      pdfauthor={},
      pdfsubject={},
      pdfkeywords={},
      pdfpagemode=None,
      bookmarksopen=true
]{hyperref}

\newcommand{\bm}{\begin{multiline}}
\newcommand{\beq}{\begin{equation}}
\newcommand{\eeq}{\end{equation}}
\newcommand{\beqs}{\begin{eqnarray}}
\newcommand{\eeqs}{\end{eqnarray}}

\begin{document}

\thispagestyle{empty}

\hfill{}

\hfill{}

\hfill{}

\vspace{32pt}

\begin{center}

\textbf{\Large Heun-type solutions of the Klein--Gordon and Dirac equations in the Garfinkle--Horowitz--Strominger dilaton black hole background}

\vspace{48pt}

\textbf{Marina-Aura Dariescu,}\footnote{E-mail: \texttt{marina@uaic.ro}}
\textbf{Ciprian Dariescu,}\footnote{E-mail: \texttt{ciprian.dariescu@uaic.ro}}
\textbf{Cristian Stelea,}\footnote{E-mail: \texttt{cristian.stelea@uaic.ro}}

\vspace*{0.2cm}

\textit{$^{1,2}$ Faculty of Physics, ``Alexandru Ioan Cuza" University}\\[0pt]
\textit{11 Blvd. Carol I, Iasi, 700506, Romania}\\[.5em]

\textit{$^3$ Research Department, Faculty of Physics, ``Alexandru Ioan Cuza" University}\\[0pt]
\textit{11 Blvd. Carol I, Iasi, 700506, Romania}\\[.5em]

\end{center}

\vspace{30pt}

\begin{abstract}
We study the Klein-Gordon and the Dirac equations in the background of the Garfinkle-Horowitz-Strominger black hole, in the Einstein frame. Using a $SO(3,1) \times U(1)-$gauge covariant approach, as an alternative to the Newman-Penrose formalism for the Dirac equation, it turns out that these solutions can be expressed in terms of Heun confluent functions and we discuss some of their properties.
\end{abstract}

\begin{flushleft}
\end{flushleft}

\baselineskip 1.5em
\setcounter{footnote}{0}

\newpage

\section{Introduction}

In recent years, black holes with electric or magnetic charge, in presence of a scalar field called {\it dilaton}, have been studied mainly in string theories. These charged black holes are solutions of the low-energy four-dimensional effective theories obtained by dimensional compactification of the heterotic string theories. Generically, the effective action of these theories describe a massless dilaton coupled to an abelian vector field \cite{Duff:1994jr}. Due to the dimensional compactification process, the dilaton is also non-minimally coupled to the Ricci scalar, the effective solution being described in the so-called string frame. However, to facilitate comparison with the standard black holes in general relativity, it is convenient to go to the so-called Einstein frame, by performing a conformal rescaling of the metric (for a review see \cite{Faraoni:1998qx}).

A remarkable black hole solution of the effective four-dimensional compactified theory was found by Gibbons and Maeda \cite{Gibbons:1982ih}, \cite{Gibbons:1987ps} and independently re-discovered in a simpler form, few years later, by Garfinkle, Horowitz and Strominger (GHS) \cite{Garfinkle:1990qj} (for a review of its properties see \cite{Horowitz:1992jp}). Even though, in terms of the string metric, the electric and magnetic black holes have very different properties, in the Einstein frame the metric doesn’t change when we go from an electrically charged to a magnetically charged black hole (this is basically due to the electromagnetic duality present in the Einstein frame. In the string frame the electromagnetic field strength is also modified by the dilaton field \cite{Casadio:1998wu}).

Using the GHS metric in the Einstein frame, the present work is devoted to a study of the Klein - Gordon and Dirac equations, which describe charged particles evolving in the Garfinkle-- Horowitz--Strominge (GHS) dilaton black hole spacetime. Within a $SO(3,1) \times U(1)-$gauge covariant approach, it turns out that the solutions can be expressed in terms of Heun confluent functions \cite{Ronveaux}, \cite{Slavyanov}.
A special attention is given to the resonant frequencies, which arise here by imposing a polynomial form of the Heun functions.
In general, the so-called quasinormal modes have a discrete spectra of complex characteristic frequencies, with the real part representing the actual frequency of the oscillation and the imaginary part representing the damping. By comparing these modes with the gravitational waves observed in the universe, one should be able to identify the presence of a GHS black hole \cite{Chen:2004zr}, \cite{Chen:2005rm}, (see also in \cite{McCarthy:2018zze} the effect of the dilaton field imprint on the gravitational waves emitted in the collision of two GHS black holes).

When the parameter related to the dilaton field goes to zero, one obtains the Klein-Gordon and Dirac equations for the usual Schwarzschild metric, which have been intensively worked out both in their original form and in different types of extensions. For instance, recently, for the Schwarzschild metric in the presence of an electromagnetic field, the Klein--Gordon and Dirac equation for massless particles have been put into a Heun-type form \cite{Birkandan:2017rdp}, \cite{Hortacsu:2011rr}. One should note Heun functions are often encountered when studying the propagation of various test fields in the background of various black holes or relativistic stars \cite{Konoplya:2002ky}- \cite{Fiziev:2009wn} and also in cosmology, in the context of extended effective field theories of inflation \cite{Ashoorioon:2018ocr}.

The method used in the present paper, while based on Cartan's formalism, is an alternative to the Newman-Penrose (NP) formalism
\cite{NP}, which is usually employed for solving Dirac equation describing fermions in the vicinity of different types of black holes \cite{Konoplya:2011qq} - \cite{Al-Badawi:2017fja}.

The structure of this paper is as follows: in the next section we present the solutions of the Klein-Gordon and Dirac equation in the background of the GHS dilatonic black hole. In section \ref{Dirmassless} we discuss the solutions of the massless Dirac equations in this background and show how to recover the expression of the Hawking temperature. The final section is dedicated to conclusions.

\section{Klein--Gordon and Dirac Equations on the GHS dilaton black hole metric}

In Einstein frame, the static and spherically symmetric GHS dilaton black hole metric is given by \cite{Garfinkle:1990qj}
\begin{equation}
ds^2=- R \, dt^2  +  \frac{dr^2}{R} + r(r-a) \left[ d \theta^2 +  \sin^2 \theta \, d \varphi^2 \right],
\label{GHS} 
\end{equation}
where
\begin{equation}
R = 1 - \frac{2M}{r} \; , \; \;
{\rm and} \; \;
a = \frac{Q^2}{M} \; ,
\end{equation}
with $M$ and $Q$ being the mass and the charge of this black hole, which has an event horizon at $r=2M$ and two singularities located at $r=0$ and $r=a$. Obviously, if the electric charge of the GHS black hole is zero, the metric (\ref{GHS}) reduces to the Schwarzschild one.

The parameter $a$ is is related to the dilaton field $\phi$ as\footnote{Note that we set the asymptotic value of the dilaton field $\phi_0=0$.}
\[
e^{-2 \phi} = 1 \mp \frac{a}{r} \, ,
\]
where the {\it minus} and {\it plus} signs are for the magnetically respectively electrically charged black holes.

Within the $SO(3,1)-$gauge covariant formulation
, we introduce the pseudo-orthonormal frame $\lbrace E_a \rbrace_{(a=\overline{1,4})}$, i.e.
\[
E_1= \sqrt{R} \, \partial_r  \; , \; \;
E_2= \frac{1}{\sqrt{r(r-a)}}  \, \partial_{\theta} \; , 
\; \; E_3 = \frac{1}{\sqrt{r(r-a)} \sin \theta}  \, \partial_{\varphi} \; , \; \;
E_4 = \frac{1}{\sqrt{R}} \, \partial_t \; ,  
\]
whose corresponding dual base is 
\[
\omega^1= \frac{1}{\sqrt{R}} \, dr\; , \; \;
\omega^2= \sqrt{r(r-a)} \,d\theta \; , 
\; \; \omega^3= \sqrt{r(r-a)} \sin \theta \,  d\varphi \; , \; \;
\omega^4= \sqrt{R} \, dt \; ,  
\]
so that the metric (\ref{GHS}) becomes the usual Minkowsky metric $ds^2 = \eta_{ab} \, \omega^a \omega^b$, with $\eta_{ab} = diag \left[ 1 , \, 1 , \, 1 , \, -1 \right]$.

Using the first Cartan's equation, 
\begin{eqnarray}
d\omega^a&=&\Gamma^a_{.[bc]}\,\omega^b\wedge \omega^c \, ,
\end{eqnarray}
with $1 \leq b<c \leq4$ and $\Gamma^a_{.[bc]} = \Gamma^a_{.bc} - \Gamma^a_{. cb}$,
we obtain the following connection one-forms $\Gamma_{ab} = \Gamma_{abc} \omega^c$, where $\Gamma_{abc} = - \Gamma_{bac}$, namely
\begin{equation}
\Gamma_{212} \, = \, \Gamma_{313} =  \frac{r-a/2}{r(r-a)} \sqrt{R} \; , \;
\Gamma_{323} \, = \, \frac{\cot \theta}{\sqrt{r(r-a)}} \; , \;
\Gamma_{414} \, = \, - \, \frac{M}{r^2 \sqrt{R}} \, 
\end{equation}

 In the pseudo-orthonormal bases (with $\eta_{44} =-1$), the fourth component of the one-form potential is:
\begin{equation}
A_4 = - \, \frac{1}{\sqrt{R}} \frac{Q}{r}  \;,
\label{em1}
\end{equation}
and it corresponds to an electric field:
\[
F_{14} = E_1 A_4 - E_4 A_1 + A_c \Gamma^c_{\; ab} - A_c \Gamma^c_{\; ba} = \frac{Q}{r^2}  \; .
\]

\subsection{The Klein-Gordon equation}

For the complex scalar field of mass $m_0$,
minimally coupled to gravity, the Klein--Gordon equation has the general $SO(3,1) \times U(1)$ gauge-covariant form
\[
\eta^{ab} \Phi_{;ab} - m_0^2 \Phi \, = \, 0 \, , 
\]
i.e.
\begin{equation}
\eta^{ab} \Phi_{|ab} - \eta^{ab} \Phi_{|c} \Gamma^c_{\; ab} \, = \, m_0^2 \Phi \,  + 2iq \, A^a \, \Phi_{|a} + q^2 A_a A^a \, \Phi \, ,
\label{KG}
\end{equation}
where
\[
\Phi_{;a} = \Phi_{|a}  - i q A_a \Phi \; ,
\]
with $\Phi_{|a} = E_a \Phi$.

The two terms in the lhs of the relation (\ref{KG}) being respectively given by
\[
\eta^{ab} \Phi_{|ab}
= R \frac{\partial^2 \Phi}{\partial r^2} + \frac{M}{r^2} \frac{\partial \Phi}{\partial r} + \frac{1}{r(r-a)} \left[ \frac{\partial^2 \Phi}{\partial \theta^2} + \frac{1}{\sin^2 \theta} \frac{\partial^2 \Phi}{\partial \varphi^2} \right] - \frac{1}{R} \frac{\partial^2 \Phi}{\partial t^2} \, 
\]
and
\[
 \eta^{ab} \Phi_{|c} \Gamma^c_{\; ab} \, = \,
 - \left[
 \frac{M}{r^2}+\frac{R(2r-a)}{r(r-a)} \right]  \frac{\partial \Phi}{\partial r} - \frac{\cot \theta}{r(r-a)} \frac{\partial \Phi}{\partial \theta} \, ,
 \]
the Klein--Gordon equation (\ref{KG}) can be cast into the following explicit form
\begin{eqnarray}
& &
r(r-a)R \frac{\partial^2 \Phi}{\partial r^2} + (2r-2M-a)  \frac{\partial \Phi}{\partial r} + \left[ \frac{\partial^2 \Phi}{\partial \theta^2} + \cot \theta \frac{\partial \Phi}{\partial \theta} + \frac{1}{\sin^2 \theta} \frac{\partial^2 \Phi}{\partial \varphi^2} \right] \nonumber \\*
& &  - \left[ r(r-a) m_0^2 + \frac{r(r-a)}{R} \left( \frac{\partial \;}{\partial t} + i \frac{qQ}{r} \right)^2 \right] \Phi =0 \; .
 \end{eqnarray}
 Using the separation of the variables with the ansatz
 \begin{equation}
 \Phi ( r , \theta , \varphi , t ) = G(r) Y_{\ell}^m ( \theta , \varphi) e^{-i \omega t} \, ,
\end{equation}
where $Y_{\ell}^m$ are the spherical harmonics, it turns out that the unknown function $G(r)$ is the solution of the differential equation
\begin{eqnarray}
& &
r(r-a)R \frac{d^2G}{dr^2} + (2r-2M-a)  \frac{dG}{d r} 
\nonumber \\*
& & -  \left[ \ell ( \ell +1) + r(r-a) m_0^2 - \frac{r-a}{r-2M} \left( \omega r -qQ \right)^2 \right] G =0 \; .
\end{eqnarray}
This equation can be solved exactly, its solutions being expressed in terms of the Confluent Heun functions \cite{Ronveaux}, \cite{Slavyanov} as:
\begin{equation}
G = e^{\frac{\alpha x}{2}} \left \lbrace C_1 \,  x^{\beta/2} HeunC \left[ \alpha , \beta , \gamma , \delta , 
\eta , x \right] + C_2 \,  x^{- \beta/2} HeunC \left[ \alpha , - \beta , \gamma , \delta , 
\eta , x \right] \right \rbrace
\label{KGHeun}
\end{equation}
with the variable
\[
x = \frac{r-2M}{a-2M}
\]
and parameters
\begin{eqnarray}
& &
\alpha = 2i (a-2M) \sqrt{\omega^2 - m_0^2} \, , \; \beta = 2i \left( 2M \omega -qQ \right) \,  , \, \gamma = 0 \, , \nonumber \\*
& &
\delta = 2 \left[ M m_0^2 - 2 M \omega^2 + qQ \omega \right] (2M-a) , \,
\eta = - \delta - \ell (\ell +1) \, . \nonumber
\end{eqnarray}

The $\alpha$ and $\beta$ parameters being purely imaginary, the radial part of the density probability is given by the square modulus of the Heun functions in (10). The two independent solutions have the generic behavior represented in the figure 1, for $r>2M>a$. The main features of the probability curve are quite nice, i.e. it satisfies all the text-book requirements imposed to a physically meaningful wave functions. In this respect, $\left| G(r) \right|^2 =1$ on the horizon and it gets a series of local decreasing maxima, finally vanishing rapidly, at the spatial infinity. If the number of these maxima was finite, the state would be bounded. Otherwise, it could asymptotically radiate.
More details on the physical phenomena related to these properties are thoroughly discussed in \cite{Vieira:2018djw}, where the authors are computing the complex values of the energy spectrum coming from the polynomial condition imposed on the Heun functions. In \cite{Vieira:2018djw} and \cite{Vieira:2016ubt}, the authors were working in coordinate bases, with 
\[
A_4^{(c)} = - \frac{Q}{r} \, ,
\]
so that $A_4^{(c)} dt = A_4 \, \omega^4$, where $A_4$ is given in (\ref{em1}).  

\begin{figure}
  \centering
  \includegraphics[width=0.45\textwidth]{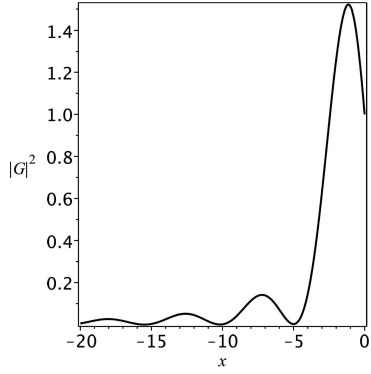} 
  \caption{The square modulus of the function $G (r)$ given in (10).} 
\label{fig}
\end{figure}

In the particular case $a=0$, corresponding to the familiar Schwarzschild black hole, the function $G$ has the same expression as in (\ref{KGHeun}), but with the variable and parameters computed for $a=0$.

\subsection{The Dirac equation}

The spinor of mass $\mu$ minimally coupled to gravity is described by the Dirac equation
\begin{equation}
\gamma^a \, \Psi_{;a} + \mu \Psi  = 0 \,
\label{Dirac}
\end{equation}
with
\[
\Psi_{;a} = \Psi_{|a} + \frac{1}{4} \, \Gamma_{bca} \,  \gamma^b  \gamma^c \Psi  - i q A_a \Psi  \; .
\]

In contrast to the Klein-Gordon case, the situation is more complicated in the case of the Dirac equation (\ref{Dirac}) and this complication is basically due to the square root $\sqrt{r(r-a)}$, which appears in the expressions of $E_2$ and $E_3$.
Thus, with the term expressing the Ricci spin-connection given by
\begin{equation}
 \frac{1}{4} \, \Gamma_{bca} \, \gamma^a \gamma^b \gamma^c =\frac{1}{2} \left[ \frac{2r-a}{r(r-a)} \sqrt{R} + \frac{M}{r^2 \sqrt{R}} \right]   \gamma^1
+ \frac{\cot \theta}{2 \sqrt{r(r-a)}} \, \gamma^2 \, \; ,
 \end{equation}
the Dirac equation becomes
\begin{eqnarray}
& &
\gamma^1 \left[ \sqrt{R} \frac{\partial \Psi}{\partial r} + \frac{2r - a -3M + aM/r}{2 \sqrt{R} r(r-a)} \Psi \right] 
+ \frac{\gamma^2}{\sqrt{r(r-a)}}  \left[ \frac{\partial \Psi}{\partial \theta} + \frac{\cot \theta}{2} \Psi \right]
\nonumber \\* & & + 
\, \frac{\gamma^3}{\sqrt{r(r-a)} \sin \theta} \,  \frac{\partial \Psi}{\partial \varphi} + \, \frac{\gamma^4}{\sqrt{R}}  \left[ \frac{\partial \Psi}{ \partial t} + iq \frac{Q}{r} \Psi \right] + \mu \Psi   = 0 \, .
\end{eqnarray}
As in the previous Klein-Gordon case, one can use the separation of the variables
\begin{equation}
\Psi = \psi ( r , \theta ) e^{i(m \varphi  - \omega t)}   \; ,
\label{psi}
\end{equation}
with the function $\psi ( r , \theta)$ defined as
\begin{equation}
\psi ( r , \theta ) = \left[ r(r-a) \sqrt{R} \right]^{-1/2}  \chi (r , \theta)  
\end{equation}
and one obtains the explicit expression of the differential equation satisfied by $\chi(r,\theta)$
\begin{eqnarray}
& &
\sqrt{R r(r-a)} \gamma^1 \, \frac{\partial \chi}{\partial r}  + \gamma^2 D_{\theta} \chi + \frac{im}{\sin \theta} \, \gamma^3 \chi \nonumber \\*
& &
+ \, i \sqrt{\frac{r(r-a)}{R}} \left( \frac{qQ}{r} - \omega \right) \gamma^4 \chi 
+ \mu \sqrt{r(r-a)} \chi = 0 \; ,
\label{chieq}
\end{eqnarray}
where
\[
D_{\theta} = \frac{\partial \;}{\partial \theta} + \frac{\cot \theta}{2} \, .
\]

Using the Weyl representation for the $\gamma^i$ matrices,
\begin{equation}
\gamma^1 = -i \beta \, \alpha^3 \; , \; \; \gamma^2  = - i \beta \alpha^1 \; , \; \;
\gamma^3 = -i \beta \alpha^2 \; , \; \; \gamma^4
= - i \beta
\, ,
\end{equation}
with
\[
\alpha^{\mu} = \left(
\begin{array}{cc}
\sigma^{\mu} & 0 \\
0 & - \sigma^{\mu} 
\end{array}
\right)  \; , \; \; \beta = \left(
\begin{array}{cc}
0 &  - {\rm I} \\
{\rm -I} & 0  
\end{array}
\right)  \; , \; \; {\rm so \; that}  \; 
\gamma^5 = \left(
\begin{array}{cc}
 {\rm I} &  0 \\
0 & {\rm -I}   
\end{array}
\right)  ,
\]
where $\sigma^{\mu}$ denote the usual Pauli matrices,
the equation (\ref{chieq}) becomes
\begin{eqnarray}
& &
\sqrt{R r(r-a)} \, \alpha^3 \, \frac{\partial \chi}{\partial r}  + \alpha^1 D_{\theta} \chi + \frac{im}{\sin \theta} \, \alpha^2 \chi \nonumber \\*
& &
+ \, i \sqrt{\frac{r(r-a)}{R}} \left( \frac{qQ}{r} - \omega \right) \chi 
+ i \mu \sqrt{r(r-a)} \beta \chi = 0 \; ,
\end{eqnarray}
and one may use again the standard procedure based on the separation of the variables. Thus, with the bi-spinor $\chi$ written in terms of two components spinors as
\begin{equation}
\chi ( r , \theta) = \left[ \begin{array}{c}
\zeta  ( r , \theta ) \\ \eta ( r , \theta )
\end{array} \right] ,
\label{chia}
\end{equation}
where 
\[
\zeta_1 = S_1 (r) T_1 (\theta ) \; , \;
\zeta_2 = S_2 (r) T_2 (\theta ) \; , \;
\eta_1 = S_2 (r) T_1 (\theta ) \; , \;
\eta_2 = S_1 (r) T_2 (\theta ) \; ,
\]
one obtains the the following system of coupled radial equations for the components $S_1$ and $S_2$, i.e.
\begin{eqnarray}
& &
S_1^{\prime} + \frac{i(qQ - \omega r)}{rR} \, S_1 + \frac{1}{\sqrt{R}} \left[ \frac{\lambda}{\sqrt{r(r-a)}} - i \mu \right] S_2 = 0 \; ,
\nonumber \\*
& &
S_2^{\prime} - \frac{i(qQ - \omega r)}{rR} \, S_2 + \frac{1}{\sqrt{R}} \left[ \frac{\lambda}{\sqrt{r(r-a)}} + i \mu \right] S_1 = 0 \,,
\label{sys}
\end{eqnarray}
if we take into account the following essential relations:
\begin{eqnarray}
& &
\left[ \frac{d \; }{d \theta} + \frac{\cot \theta}{2} + \frac{m}{ \sin \theta} \right] T_2 = \lambda T_1  \; , \nonumber \\*
& &
\left[ \frac{d \; }{d \theta} + \frac{\cot \theta}{2} - \frac{m}{\sin \theta} \right] T_1 =  - \, \lambda T_2 \; .
\end{eqnarray}

Thus, the angular parts $T_A$, with $A= 1,2$, are satisfying the decoupled equations
\begin{equation}
\frac{d^2T_A}{d \theta^2} + \cot \theta \frac{dT_A}{d \theta} - \left[ \frac{( \cos \theta \mp 2m)^2}{4 \sin^2 \theta} - \lambda^2 + \frac{1}{2} \right] T_A = 0  \, ,
\end{equation}
with the solutions given by the spin-weighted spherical harmonics \cite{Wang:2017fie}, for
$\lambda = \ell + 1/2$.

As for the radial equations, we employ the auxiliary function method and consider $S_1$ and $S_2$ as being
\begin{eqnarray}
& &
S_1 = e^{i \omega r} \left( \frac{r}{2M}-1 \right)^{2i \omega M - i qQ} \Sigma_1 (r) \; , \nonumber \\*
& & S_2 = e^{- i \omega r} \left( \frac{r}{2M}-1 \right)^{- 2i \omega M + i qQ} \Sigma_2 (r) \, ,
\end{eqnarray}
so that the system (\ref{sys}) leads to the following simpler equations for the unknown functions $\Sigma_A$:
\begin{eqnarray}
& &
\Sigma_1^{\prime} + 
\left( \frac{r}{2M}-1 \right)^{-4i \omega M +2 i qQ}
\frac{e^{-2i \omega r}}{\sqrt{R}}
\left[ \frac{\lambda}{\sqrt{r(r-a)}} - i \mu \right] \Sigma_2 = 0 \; ,
\nonumber \\*
& &
\Sigma_2^{\prime} + 
\left( \frac{r}{2M}-1 \right)^{4i \omega M -2 i qQ}
\frac{e^{2i \omega r}}{\sqrt{R}}
\left[ \frac{\lambda}{\sqrt{r(r-a)}} + i \mu \right] \Sigma_1 = 0 \; .
\end{eqnarray}
The differential equation for $\Sigma_1$, i.e.
\begin{eqnarray}
& & \Sigma_1^{\prime \prime} +
\left[ \frac{4i \omega r-4iqQ+1}{2(r-2M)} + \frac{1}{2(r-a)} +
\frac{i \mu (2r-a)}{2 \sqrt{r(r-a)} \left[ \lambda - i \mu \sqrt{r(r-a)} \right]}
\right] \Sigma_1^{\prime} 
\nonumber \\*
& & - \, \frac{r}{r-2M} \left[ \frac{\lambda^2}{r(r-a)} + \mu^2 \right] \Sigma_1 = 0 \; ,
\label{sigma1}
\end{eqnarray}
can not be analitically solved. 
Numerically, using Mathematica \cite{math}, with the initial conditions
\[
\Sigma_1 ( 2M_+ ) = 0 \; , \;
\Sigma_1^{\prime} ( 2 M_+ ) = 1 \; ,
\]
the absolute value of the radial part of $| \psi |^2$ given in (\ref{psi}), namely
\begin{equation}
F(r) = \frac{1}{r(r-a) \sqrt{R}} \left| S_1 \right|^2 \, ,
\label{Fr}
\end{equation}
is represented in the figure \ref{fig}, for $r>2M>a$.

This is describing the fermionic ground state in the outer region, with just one maximum (as it should) and exponentially vanishing at infinity. We haven't analyzed the corresponding modes located within the black hole, since, in the limit $a \to 0$, the area of the sphere $r=a$ is zero so that this surface is singular. Once $Q$ increases,
the singular surface moves towards the event horizon $r=2M$, and one has to solve the problems related to the physically meaningful boundary conditions.

\begin{figure}
  \centering
  \includegraphics[width=0.45\textwidth]{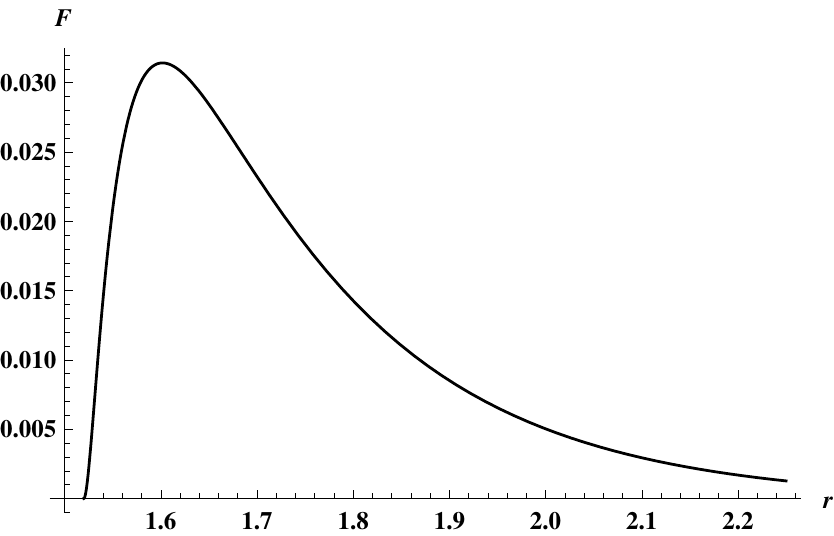} 
  \caption{The function (\ref{Fr}), for $a=1$ and $2M =1.5$.} 
\label{fig}
\end{figure}

If one imposes $\lambda \ll \mu r$ and performs a series expansion of the last term multiplying $\Sigma_1^{\prime}$ in (\ref{sigma1}), to first order in $a/r$, this last term can be approximated to $1/(r-a)$ and the solution is given by the Heun confluent functions \cite{Ronveaux}, \cite{Slavyanov} as
\begin{eqnarray}
\Sigma_1 & = & C_1 \,  e^{- \mu r} HeunC \left[ \alpha , \, \beta , \, \gamma , \, \delta , \, \eta , \, \frac{r-2M}{a-2M} \right] 
\nonumber \\* & &
+ \, C_2 \, e^{- \mu r} \left( \frac{r}{2M}-1 \right)^{-2i \omega M + 2i qQ +1/2} HeunC \left[ \alpha , \, - \beta , \, \gamma , \, \delta , \, \eta , \, \frac{r-2M}{a-2M}
\right],
\label{sigma1ex}
\end{eqnarray}
with the parameters
\begin{eqnarray}
& &
\alpha = 2 \mu (2M -a) , \; \beta = 2i ( \omega M - qQ) - \frac{1}{2} , \, \gamma = - \frac{3}{2} , \,
\delta = 2 \mu^2 M (2M-a)  \, \nonumber \\*
& &
\eta = - \delta + \frac{i( \omega M -qQ)}{2} + \frac{5}{8} - \lambda^2 \, .
\label{par}
\end{eqnarray}

Near the exterior event horizon, $r \to 2M$ i.e. $x \to 0$, the Heun confluent functions in (\ref{sigma1ex}) have a polynomial form if their parameters are satisfying the condition \cite{Ronveaux}, \cite{Slavyanov}:
\begin{equation}
\frac{\delta}{\alpha} = - \left[ n+1 + \frac{\beta+\gamma}{2} \right] .
\label{delta}
\end{equation}
By replacing the expressions (\ref{par}) in (\ref{delta}), one gets the relation
\begin{equation}
i ( \omega M - qQ) + \mu M = - n \, ,
\end{equation}
for the Heun function multiplied by $C_1$ and
\begin{equation}
i ( \omega M - qQ) - \mu M = n + \frac {1}{2} \, ,
\end{equation}
for the one multiplied by $C_2$.
These relations are
pointing out a quantized part of the imaginary part of $\omega$, which corresponds to resonant frequencies \cite{Vieira:2018djw}.

\section{The Massless Case}
\label{Dirmassless}

The Dirac equation has been worked out for several physically important metrics, mainly using the NP formalism \cite{Kraniotis:2018zmh} and some of the solutions, especially in the massless case, have been expressed in terms of Heun confluent functions \cite{Birkandan:2017rdp}.

In view of the analysis developed in the previous section, the massless and chargeless fermions are described by the radial equations coming from the system (\ref{sys}), namely
\begin{eqnarray}
& &
S_1^{\prime} - \frac{i\omega}{R} \, S_1 + \frac{\lambda}{\sqrt{R r(r-a)}} S_2 = 0 \; ,
\nonumber \\*
& &
S_2^{\prime} + \frac{i\omega}{R} \, S_2 + \frac{\lambda}{\sqrt{R r(r-a)}} S_1 = 0 \; ,
\label{sys2}
\end{eqnarray}
with
\begin{eqnarray}
& &
S_1 = e^{i \omega r} \left( \frac{r}{2M}-1 \right)^{2i \omega M} \Sigma_1 (r) \; , \nonumber \\*
& & S_2 = e^{- i \omega r} \left( \frac{r}{2M}-1 \right)^{- 2i \omega M } \Sigma_2 (r) \, .
\label{Si}
\end{eqnarray}
Thus, the system (\ref{sys2}) turns into the simpler form
\begin{eqnarray}
& &
\Sigma_1^{\prime} + 
e^{-2i \omega r}
\left( \frac{r}{2M}-1 \right)^{-4i \omega M}
\frac{\lambda}{\sqrt{R r(r-a)}}
\Sigma_2 = 0 \; ,
\nonumber \\*
& &
\Sigma_2^{\prime} + 
e^{2i \omega r}
\left( \frac{r}{2M}-1 \right)^{4i \omega M}
\frac{\lambda}{\sqrt{R r(r-a)}}
\Sigma_1 = 0 \; ,
\end{eqnarray}
which leads to the following differential equation for $\Sigma_1$,
\begin{equation}
\Sigma_1^{\prime \prime} +
\left[ \frac{4i \omega r+1}{2(r-2M)} + \frac{1}{2(r-a)} \right] \Sigma_1^{\prime} - \frac{\lambda^2}{(r-2M)(r-a)} \Sigma_1 = 0
\end{equation}
and similarly for $\Sigma_2$.
The solution of this equation is expressed in terms of Heun confluent functions as
\begin{eqnarray}
\Sigma_1 = e^{-2i \omega r} \left \lbrace C_1 \,   HeunC \left[ \alpha , \beta , \gamma , \delta , \eta , \, x \right] 
+ \, C_2 \,  x^{- \beta}  HeunC \left[ \alpha , - \beta , \gamma , \delta , \eta , \,  x 
\right] \right \rbrace  
\label{sigma1m}
\end{eqnarray}
where the variable is
\[
x = \frac{r-2M}{a-2M} 
\]
and the corresponding parameters are:
\begin{eqnarray}
& &
\alpha = 2i \omega (2M -a) , \; \beta = 4i \omega M - \frac{1}{2} , \, \gamma = - \frac{1}{2} , \,
\delta = i \omega ( 4 i \omega M +1)  (2M-a) \, \nonumber \\*
& &
\eta = - \delta - \frac{i \omega a}{2} + \frac{3}{8} - \lambda^2 \, .
\label{par2}
\end{eqnarray}

The solutions to Heun's confluent equations are computed as power series expansions around the regular singular point $x=0$, i.e. $r=2M$. The series converges for $r < a$ (the second regular singularity) and the analytic continuation is obtained by expanding the solution around the regular singularity $r=a$, and overlapping the series. 

For large $x$ values, one may use the formula \cite{Ronveaux}, \cite{Vieira:2018djw}
\begin{eqnarray}
& &
HeunC \left[ \alpha , \, \beta , \, \gamma , \, \delta , \, \eta , \, x \right]
\approx D_1 x^{- \left[ \frac{\beta+\gamma+2}{2} + \frac{\delta}{\alpha} \right]}
+ D_2 e^{- \alpha x} x^{- \left[ \frac{\beta+\gamma+2}{2} - \frac{\delta}{\alpha} \right]}  \nonumber \\*
& & = e^{- \frac{\alpha x}{2}}   x^{- \frac{\beta+\gamma+2}{2}} \left \lbrace D_1 e^{\frac{\alpha x}{2}}   x^{- \frac{\delta}{\alpha}} + 
D_2 e^{- \frac{\alpha x}{2}} x^{\frac{\delta}{\alpha}} 
\right \rbrace \nonumber \\*
& &
= D e^{- \frac{\alpha x}{2}}   x^{- \frac{\beta+\gamma+2}{2}} \sin \left[
- \frac{i \alpha x}{2} + \frac{i \delta}{\alpha} \ln x + \sigma \right] ,
\end{eqnarray}
where $D$ is an arbitrary constant and $\sigma$ is the phase shift.
With the parameters given in (\ref{par2}), the component $S_1$ from (\ref{Si}) gets the asymptotic form 
\begin{equation}
S_1 \approx \frac{D}{\sqrt{r}} \sin \left[ \omega r + 2 \omega M \ln r - \frac{i}{2} \ln r + \sigma \right]
\end{equation}
which, for large $r$ values, behaves like
\begin{equation}
S_1 \sim \exp \left[  i \left( \omega r + 2 \omega M \ln r + \sigma \right) \right] .
\end{equation}

The necessary condition for a polynomial form of the confluent Heun functions (\ref{delta}), leads to the following quantized imaginary quasispectrum 
\begin{equation}
\omega =   i \, \frac{(n+1)}{4M}
\, .
\label{spec}
\end{equation}

In order to study the radiation emitted by the GHS black hole, one has to take the radial solution near the exterior event horizon, $r \to r_h = 2M$. For $x \to 0$, the Heun functions can be approximated to $1$ and the (radial) components of $\Psi$ defined in (\ref{psi}) and (\ref{chia}), with $S_A$ given in (\ref{Si}) can be written as
\begin{equation}
\Psi_1 \approx e^{-i \omega t} e^{-i \omega r} \frac{1}{\sqrt{r(r-a)}} \left \lbrace C_1 (r-2M)^{2i \omega M - \frac{1}{4}} + C_2 
(r-2M)^{- 2i \omega M + \frac{1}{4}} \right \rbrace
\end{equation}
and
\begin{equation}
\Psi_2 \approx e^{-i \omega t} e^{i \omega r} \frac{1}{\sqrt{r(r-a)}} \left \lbrace C_1 (r-2M)^{-2i \omega M - \frac{1}{4}} + C_2 
(r-2M)^{2i \omega M + \frac{1}{4}} \right \rbrace ,
\end{equation}
pointing out the {\it in} and {\it out} modes
\begin{eqnarray}
\Psi_{in} & \sim & e^{-i \omega t}  (r-2M)^{-2i \omega M + \frac{1}{4}} \; , \nonumber \\*
\Psi_{out} & \sim & e^{-i \omega t}  (r-2M)^{2i \omega M - \frac{1}{4}} 
 \, .
\end{eqnarray}
By definition, the component $\psi_{out}$ should asymptotically have the form
\begin{equation}
\Psi_{out} \sim ( r - r_h)^{\frac{i}{2 \kappa_h} ( \omega - \omega_h)}
\end{equation}
so that the relative scattering probability at the exterior event horizon surface is given by
\[
\Gamma = \left| \frac{ \Psi_{out} ( r>2M)}{\Psi_{out} (r<2M)} \right|^2 = \exp \left[ - \frac{2 \pi}{\kappa_h} ( \omega - \omega_h) \right] .
\]
Inspecting the above relations, it yields the well-known results: $\kappa_h = 1/(4M)$,
\[
\Gamma = e^{-8 \pi M \omega} 
\]
and the mean number of emitted particles
\[
N = \frac{\Gamma}{1- \Gamma} = \frac{1}{e^{\frac{\omega}{T_h} }-1}  \, ,
\]
where $T_h = 1/(8 \pi M)$ is the Hawking temperature.

\section{Conclusions}

In the present paper, we have used the free of coordinates formalism to write down both the Klein--Gordon and the Dirac equation, in their $SO(3,1) \times U(1)$ expression, for the GHS metric (\ref{GHS}).

Unlike the case for bosons, it turns out that, for the charged massive fermions interacting with the GHS dilaton black hole, the radial equation (\ref{sigma1}) does not have an analytic solution.
However, to first order in $a/r$, the corresponding equations are satisfied by the Heun confluent functions (\ref{sigma1ex}).

In the massless case, the Dirac equation can  be analytically solved and the derived solution, given by (\ref{sigma1m}), is valid for the whole space, which includes not only the near-horizon region, but also the far away from the black hole region. Once the relation (\ref{delta}) among the Heun function's parameters is imposed, the confluent Heun functions can be cast into a polynomial form and the energy spectrum is given by the imaginary quantized expression (\ref{spec}).

Finally, by identifying the {\it out} modes near the event horizon, we identified the Hawking black body radiation and the expected Hawking temperature is correctly recovered.

\begin{flushleft}
\begin{Large}
{\bf Acknowledgement}
\end{Large}
\end{flushleft}
The authors are most grateful to the anonymous referee for
insightful suggestions which have been of a real help in improving the original form of the manuscript.
The authors declare that there is no conflict of interest regarding the publication of this paper.
This work was supported by a grant of Ministery of Research and Innovation, CNCS - UEFISCDI, project number PN-III-P4-ID-PCE-2016-0131, within PNCDI III.

\end{document}